\newcommand{\nc}{\newcommand}
\nc{\kms}{\,{km\,s$^{-1}$}}
\nc{\sgra}{Sgr\ts A}
\nc{\sgrastar}{Sgr\ts$\rm {A}^{*}$}
\nc{\sgraeast}{Sgr\ts A\ts East}
\nc{\sgrawest}{Sgr\ts A\ts West}
\nc{\sgracomp}{Sgr\ts A\ts complex}
\nc{\as}{\arcsec}
\nc{\HI}{H\,{\sc i}}
\nc{\HII}{H\,{\sc ii}}
\nc{\CI}{C\,{\sc i}}
\nc{\CII}{C\,{\sc i}}
\nc{\hto}{H$_{2}$O}
\nc{\htmo}{H$_{2}^{16}$O}
\nc{\htio}{H$_{2}^{18}$O}
\nc{\am}{\arcmin}
\nc{\ciso}{C$^{18}$O}
\nc{\hctn}{HC$_{3}$N}
\nc{\ot}{O$_{2}$}
\nc{\nht}{NH$_{3}$}
\nc{\htco}{H$_{2}$CO}
\nc{\htre}{H$_{3}$O${^+}$}
\nc{\ohs}{OH-Streamer}
\nc{\s}{{$\mathrm{M}_{\odot} $}}
\nc{\snrs}{SNR (G359.92$-$0.09)}

\documentclass{iau}
\usepackage{graphicx}

\title[Hydroxyl, water, ammonia, carbon monoxide and neutral carbon towards the Sgr A complex] 
{Hydroxyl, water, ammonia, carbon monoxide and neutral carbon towards the Sgr A complex}

\author[Karlsson et al.]   
{R. Karlsson$^1$, Aa. Sandqvist$^1$, \AA{}. Hjalmarson$^2$, A. Winnberg$^2$, K. Fathi$^1$, 
U. Frisk$^3$ and M. Olberg$^2$}

\affiliation{$^1$ Stockholm Observatory, Department of Astronomy, Stockholm University, AlbaNova 
University Center, SE-106 91 Stockholm, Sweden
$^2$ Onsala Space Observatory, Chalmers University of Technology, SE-439 92 Onsala, 
Sweden
$^3$ Omnisys Instruments AB, Solna Strandv\"{a}g 78, SE-171 54 Solna, Sweden}

\pubyear{2013}
\volume{303}  
\pagerange{--}
\jname{The GC: Feeding and Feedback in a Normal Galactic Nucleus}
\editors{Editors}
\begin{document}

\maketitle

\begin{abstract}
We observed OH, \hto, \nht, \ciso, and \CI\ towards the +50 \kms\ cloud  (M--0.02--0.07), the CND and the +20 \kms\ (M--0.13--0.08) cloud in the Sgr A complex with the VLA, Odin and SEST. Strong OH absorption, \hto\ emission and absorption lines were seen at all three positions. Strong \ciso\ emissions were seen towards the +50 and +20 \kms\ clouds. The CND is rich in \hto\ and OH, and these abundances are considerably higher than in the surrounding clouds, indicating that shocks, star formation and clump collisions prevail in those objects. A comparison with the literature reveals that it is likely that PDR chemistry including grain surface reactions, and perhaps also the influences of shocks has led to the observed abundances of the observed molecular species studied here. In the redward high-velocity line wings of both the +50 and +20 km/s clouds and the CND, the very high \hto\ abundances are suggested to be caused by the combined action of shock desorption from icy grain mantles and high-temperature, gas-phase shock chemistry. Only three of the molecules are briefly discussed here. For OH and \hto\ three of the nine observed positions are shown, while a map of the \ciso\ emission is provided. An extensive paper was  recently published with Open Access (Karlsson et al. 2013).

\end{abstract}

\firstsection 
\section{Introduction}
The Karl G. Jansky Very Large Array (VLA), the Odin satellite and the Swedish ESO Submillimetre Telescope (SEST) were used to observe OH absorption (1665 and 1667 MHz), \hto\ emission and absorption ($1_{10} - 1_{01}$, 557 GHz), and \ciso\ emission ($J=2-1$, 220 GHz). A comparison of the abundances of the three molecules may serve as a useful tool to probe the chemistry of interstellar molecular clouds.

\section{Results}
The OH and \hto\ line profiles are shown in Fig. 1. The OH abundance appears to be enhanced by an order of magnitude in the CND cloud core and is very likely also enhanced in the +50 and +20 \kms\ cloud cores, compared to the abundances estimated in the line-of-sight spiral arm features. The \hto\ abundance is markedly enhanced in the front sides of the Sgr A molecular cloud cores, $(2-7) \times 10^{-8}$, as observed in absorption and being highest in the CND. A similar abundance enhancement is seen in OH. An unusually high OH/\hto\ ratio is also found in the clouds at the GC. The likely explanation is PDR chemistry including grain surface reactions, and perhaps also influences of shocks and turbulence in clouds. In the redward high-velocity line wings of the +50 and +20 \kms\ clouds and the CND, very high \hto\ abundances are found, $(1 - 6) \times 10^{-6}$, and suggested to be caused by the combined action of shock desorption from icy grain mantles and high temperature gas phase shock chemistry. In the +50 \kms\ cloud, three large clumps (I, II and III), were identified in \ciso\ emission, see Fig. 2. The surfaces of the clumps facing Sgr A East appear to be associated with 1720 MHz OH shock-excited masers. An area of depleted \ciso\ emission was found, at a location near the \snrs\ which is marked by an arc in Fig. 2. At the intersection of the \snrs\ and Sgr A East, five 1720 MHz OH masers are seen in the figure.

\begin{figure}
\begin{center}
\includegraphics[width=.3\textwidth]{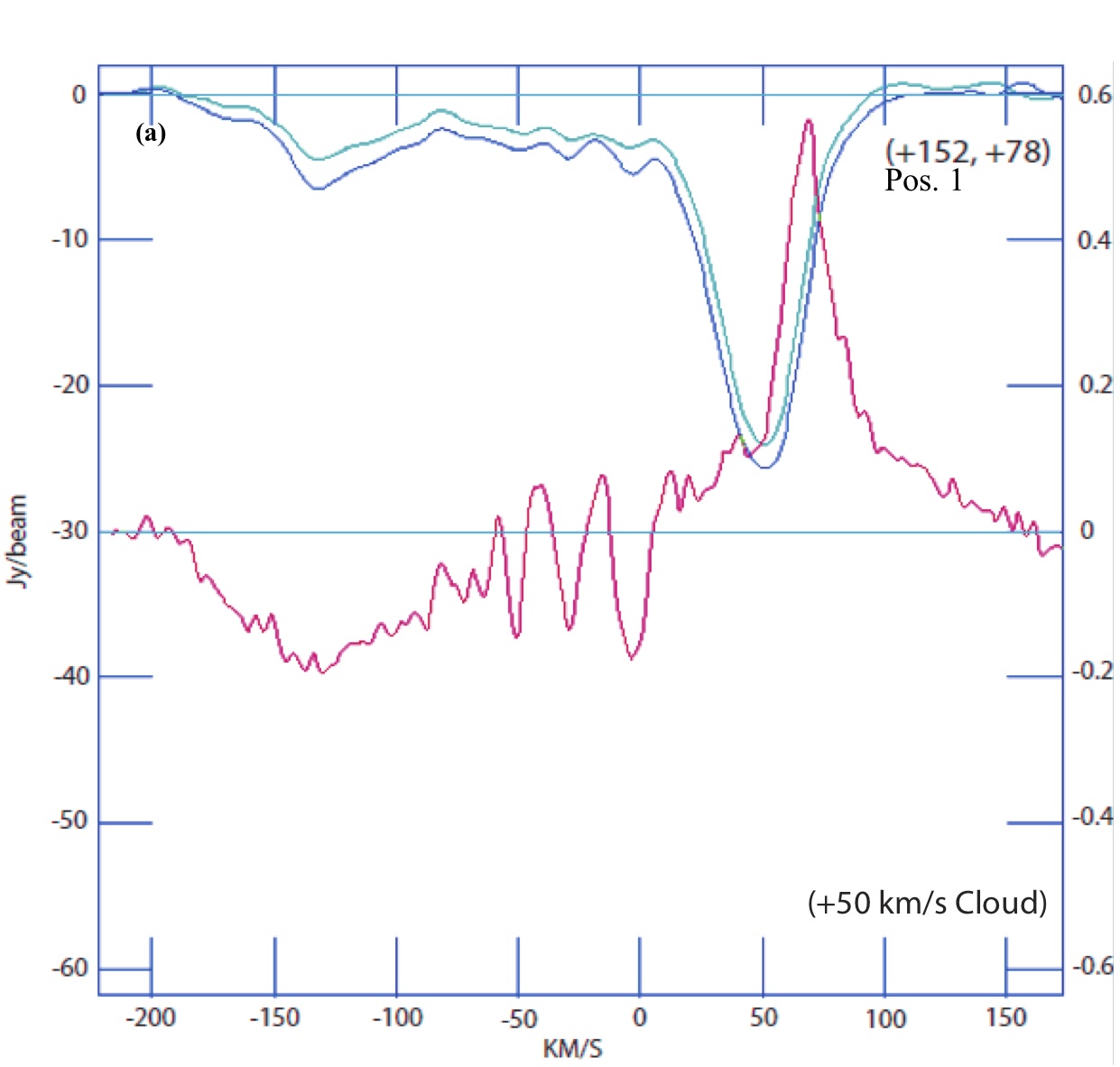}
\includegraphics[width=.3\textwidth]{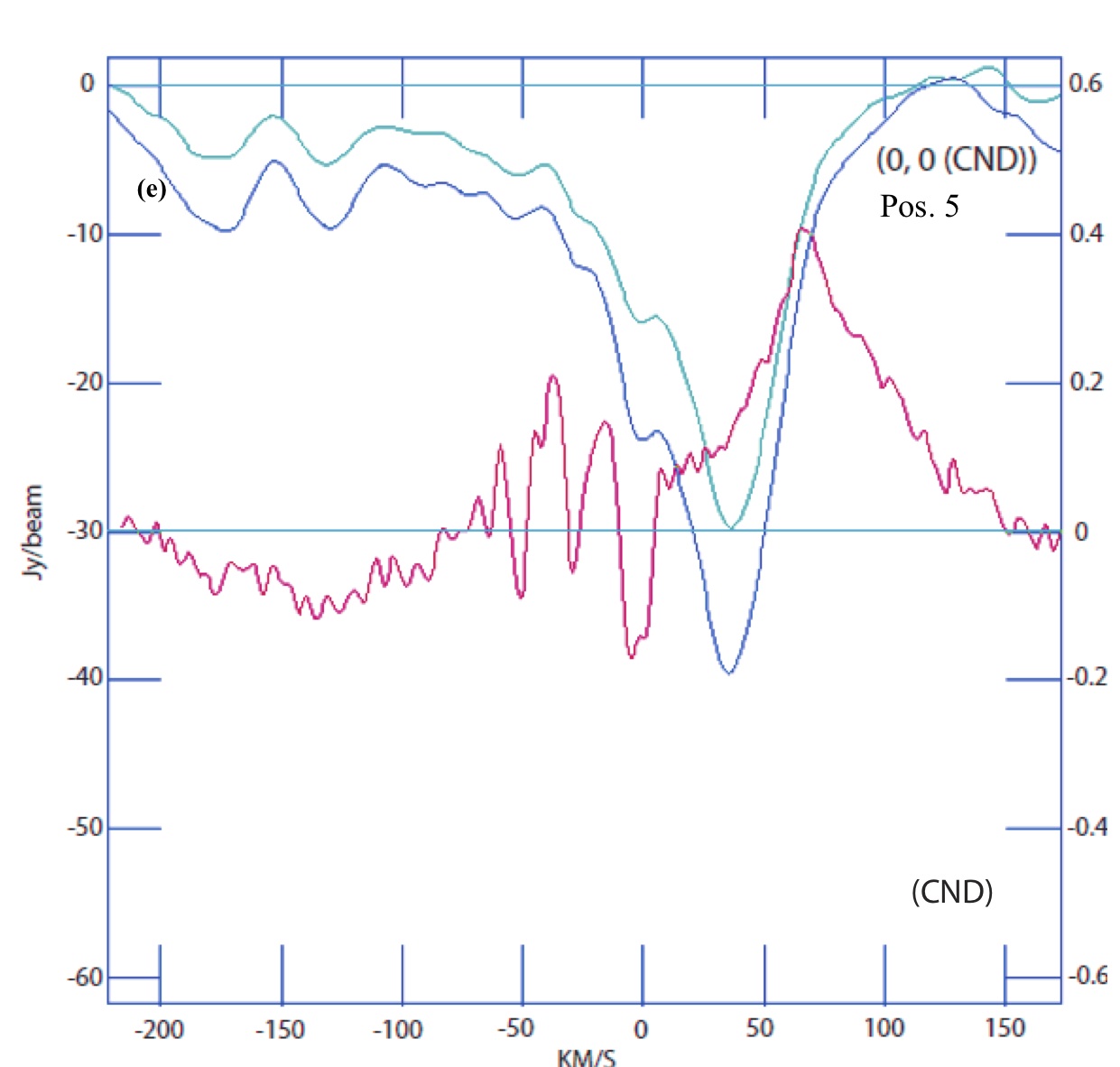}
\includegraphics[width=.3\textwidth]{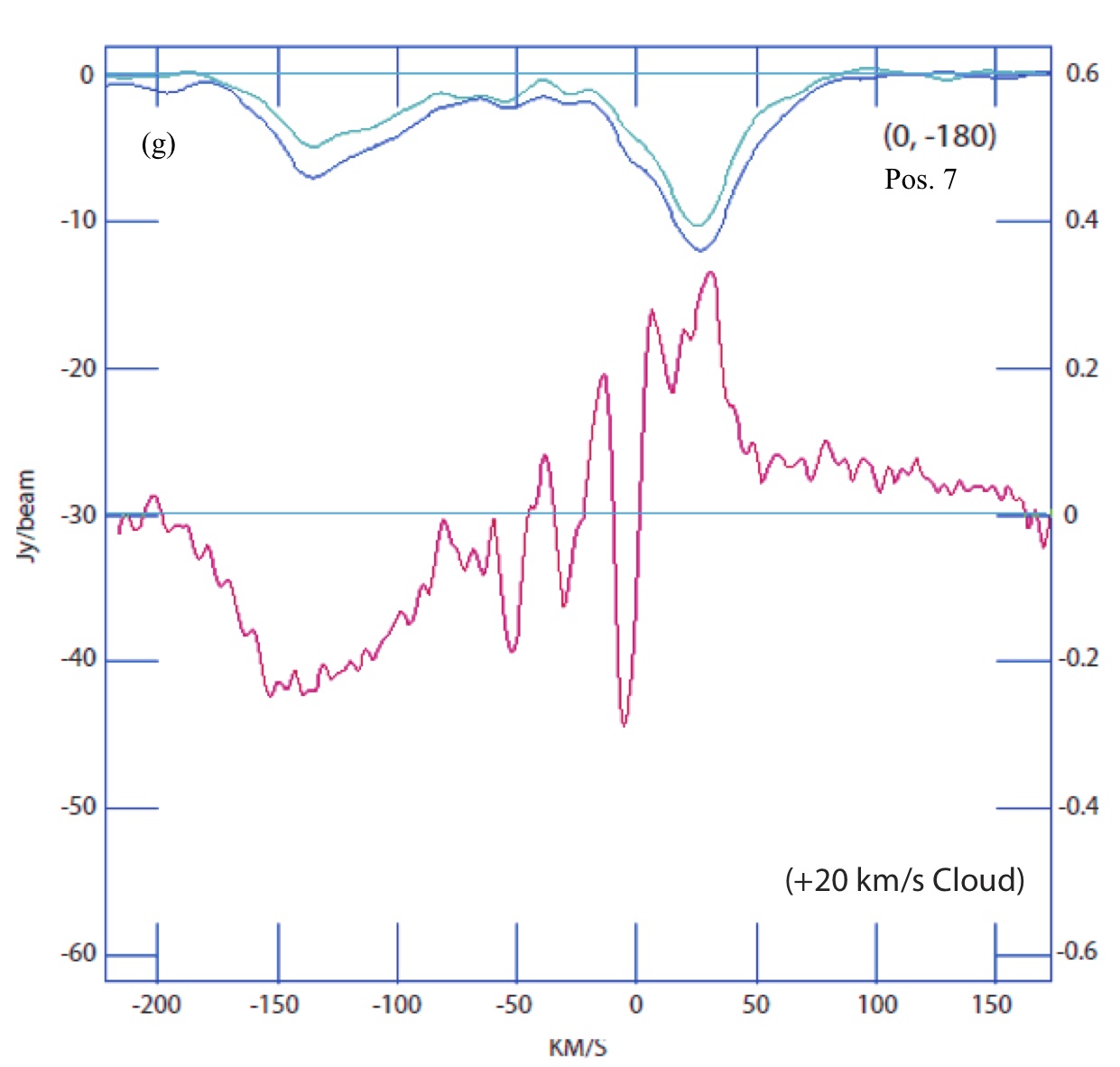}
\caption{1665 and 1667 MHz OH and \hto\ line profiles towards positions 1, 5 and 7 in Fig. 2. The upper line profiles refer to the 1665 MHz OH line, and the deeper absorption profiles belong to the 1667 MHz OH line observed with the VLA. The profiles in the middle of the diagrams are the \hto\ profiles observed with the Odin satellite.}
\label{k2fig2panel}
\end{center}
\end{figure}

\begin{figure}
  \begin{minipage}[c]{0.6\textwidth}
    \includegraphics[width=\textwidth]{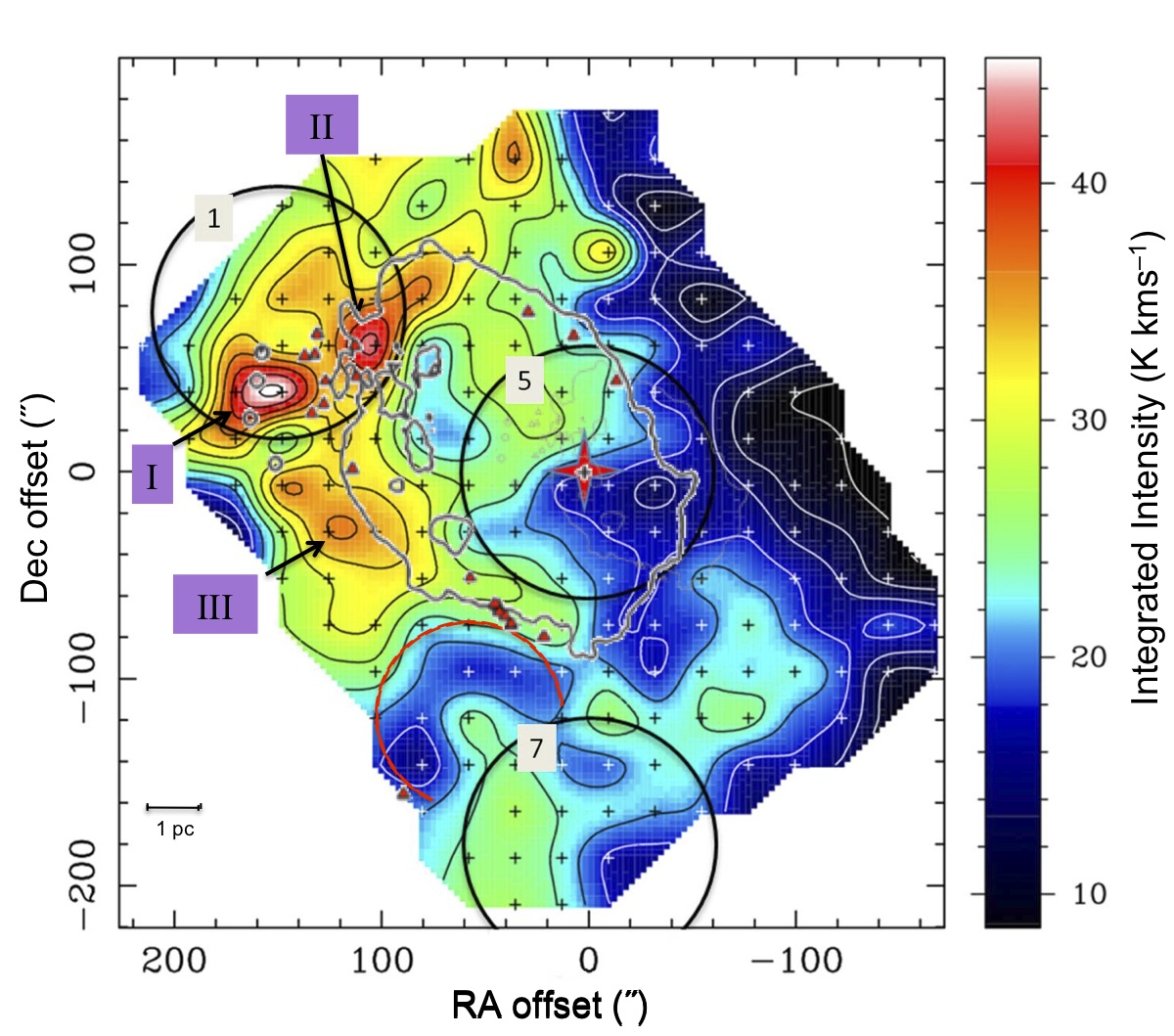}
  \end{minipage}\hfill
  \begin{minipage}[c]{0.38\textwidth}
    \caption{Total integrated intensity ($-$200 to 200 \kms) of antenna temperature of SEST observations of \ciso\ $J=2-1$ towards the Sgr A complex.  Sgr A* is marked with a star, and the large circles mark the Odin beam positions in the +50 \kms\ cloud (1), the CND (5), and the +20 \kms\ cloud (7). The tiny triangles indicate positions of 1720 MHz OH SNR masers observed by Yusef-Zadeh et al. (\cite{yus96}, \cite{yus01}), Karlsson et al. (\cite{kar03}), and Sjouwerman \& Pihlstr\" om (\cite{sjo08}), and the small circles indicate the positions of the four Compact \HII\ regions (Ekers et al. \cite{eke83}). The grey contour line delineates the Sgr A East 18-cm continuum emission at 0.15 Jy/beam (Karlsson et al. in preparation). The HPBW is 24 arcseconds.} 
  \end{minipage}
\end{figure}

\end{document}